\documentstyle[aps,epsf,psfig]{revtex} 

\newcommand{\be}[1]{\begin{equation}\label{#1}}
\newcommand{\ee}{\end{equation}}
\newcommand{\dpi}{\delta_{2\pi}}
\renewcommand{\vec}[1]{{\bf #1}}
\newcommand{\vecgr}[1]{\mbox{\boldmath{$ #1$}}}
\newcommand{\Tr}{\mbox{Tr}}
\newcommand{\tU}{\tilde{U}}
\newcommand{\smax}{\mbox{\scriptsize max}}
\newcommand{\N}{\mbox{I}\!\mbox{N}}

\newcommand{\sN}{\mbox{\scriptsize I}\!\mbox{\scriptsize N}}
\newcommand{\et}{{\em et al }}

\begin{document}
\title{
Periodic orbit action correlations in the Baker map 
}
\author{ Gregor Tanner} 
\address{
Basic Research Institute in the Mathematical Sciences,\\
Hewlett-Packard Laboratories Bristol,\\ 
Filton Road, Stoke Gifford, Bristol BS12 6QZ, UK\\
and\\
School of Mathematical Sciences
\footnote{Present address}
\footnote{e-mail: gregor.tanner@nottingham.ac.uk}\\
Division of Theoretical Mechanics\\
University of Nottingham\\
University Park, Nottingham NG7 2RD, UK
}

\maketitle

\begin{abstract}
Periodic orbit action correlations are
studied for the piecewise linear, area-preserving Baker map. 
Semiclassical periodic orbit formulae together with universal spectral
statistics in the corresponding quantum Baker map suggest the existence of 
universal periodic orbit correlations. 
The calculation of periodic orbit sums for the Baker map 
can be performed with the help of a Perron-Frobenius type operator.
This makes it possible to study periodic orbit correlations for 
orbits with period up to 500 iterations of the map. 
Periodic orbit correlations are found to agree 
quantitatively with the predictions from random matrix theory 
up to a critical length determined by the semiclassical error.
Exponentially increasing terms dominate the correlations for longer
orbits which are due to the violation of unitarity in the semiclassical
approximation. 
\end{abstract}

\section{Introduction}
\label{sec:sec1}
Numerical evidence suggests that eigenvalue spectra of 
individual low dimensional quantum systems show correlations which are solely 
determined by the underlying classical dynamics, symmetries and Planck's 
constant $\hbar$ (see e.g.\ Bohigas \et 1984, Berry 1987, 
Bohigas 1991). Especially quantum systems exhibiting fully chaotic or
completely regular dynamics in the classical limit show spectral fluctuations 
coinciding with random matrix theory (RMT). Semiclassical trace formulae 
provide a link between the set of eigenenergies and the set of all 
periodic orbits in the classical system and have proven to be important in
understanding universality in spectral statistics and deviations thereof 
(Hannay and Ozorio de Almeida 1984, Berry 1985, Berry 1988, Aurich and 
Steiner 1995, Bogomolny and Keating 1996a). A complete description of 
spectral properties in terms of the underlying classical dynamics of the 
system is, however, still missing.

Universal spectral statistics is intimately connected to correlations in 
periodic orbit length or actions distributions. The
validity of the random matrix conjecture implies generalised universal 
periodic action correlation for systems for which the trace formula is exact
(Argaman et al 1993). Keating (1993) applied this idea to the 
Riemann zeta function whose non-trivial zeros are conjectured to follow random 
matrix statistics. Here, the Hardy - Littlewood conjecture provides 
an explicit expression for the pair - correlations of prime numbers, (the
analogue of periodic orbits in the Riemann case). Expressions for the 
two-point correlation function of Riemann zeros can be obtained making use of 
the conjectured prime number correlations which
coincide asymptotically with RMT results for an ensemble of Gaussian random 
matrices invariant under unitary transformation (GUE). A generalisation of 
these arguments to n-point correlation functions for Riemann-zeros has 
been given by Bogomolny and Keating (1995, 1996). 

Far less is known about action correlations of periodic orbits for generic 
chaotic systems and even numerical evidence for the existence of these 
universal periodic orbit correlations is weak.  Argaman \et (1993) present 
numerical results which agree qualitatively with the RMT-prediction presented 
in the same paper; Cohen \et (1998) find correlations in the length 
spectrum of periodic orbits in the Stadium billiard which indicate RMT-like
correlations but a quantitative analysis can not compete with the 
accuracy reached for spectral eigenvalue statistics. Other groups report 
complete failure in their search for any kind of action correlations 
(Aurich 1998) or find action correlations which do not coincide with the 
RMT - prediction by Argaman \et (1993), see Sano (1999); 
in addition there are yet no arguments other than from the 
duality between quantum eigenvalues and classical periodic orbits which 
indicate why universal classical correlations should exist. 
Cohen \et (1998) present a variety of ideas based on 
identifying relationships within various subgroups of orbits in the Sinai
billiard. The results are, however, still of speculative nature.\\ 

In this paper I will study quantum and semiclassical spectra and discuss
necessary conditions for the existence of universal periodic orbit action 
correlations. I will focus in particular on 
the classical and quantum Baker map. It will become evident 
that the periodic orbit correlations deviate from the universal behavior for
long orbits due to the violation of quantum unitarity in the semiclassical 
approximations. Correlations following the RMT-prediction do, however, exist
for short periodic orbits which can not be explained by classical sum rules 
alone. Properties of periodic orbits can be studied in detail for the Baker 
map by writing periodic orbit sums in terms of a suitable classical 
Perron-Frobenius operator. Exponentially increasing terms in the large action 
limit can be regularised by imposing unitarity onto a semiclassical 
quantization as proposed by Bogomolny and Keating (1996a).

\section{Periodic Orbit Correlation Functions for Quantum Maps}
\label{sec:sec2}
In the following I will limit the discussion to quantum maps.
A quantum map acts on a finite dimensional Hilbert space and the 
quantum dynamics is governed by the
equation
\be{map} \psi_{n+1} = \vec{U} \psi_n \ee
with $ \vec{U}$ being a unitary matrix of dimension $N$. We assume that the 
map has a well defined classical limit for $N \to \infty$ described by a 
discrete dynamical map. $\psi_n$ is the discretised $N$ - dimensional wave 
vector and $N$ is equivalent to the inverse of Planck's constant $h$, i.e.\ 
$N \sim 1/h$. For possible quantization procedures 
of classical maps see e.g.\ Hannay and Berry (1980) for the cat map, Balazs 
and Voros (1989) and Saraceno and Voros (1994) for the Baker map or Bogomolny 
(1992), Doron and Smilansky (1991), and Prosen (1994, 1995) for semiclassical 
and quantum Poincar\'e maps.  The spectral density of eigenphases $\theta_i$ 
of $\vec{U}$ can be written in the form
\be{density}
d(\theta,N) = \sum_{i=1}^N \dpi(\theta - \theta_i) 
= \frac{N}{2\pi} + \frac{1}{\pi}{Re}\sum_{n=1}^{\infty} 
\Tr \vec{U}^n e^{-i n\theta} 
\ee
with $\overline{d} = N/2\pi$ being the mean density of eigenphases in the 
interval $[0,2\pi]$ and $\dpi$ denotes the periodically continued delta 
function with period $2 \pi$. The spectral measure of interest here is the 
two-point correlation function defined as 
\be{two-point1}
R_2(x,N) = \frac{1}{\overline{d}^2} <d(\theta) d(\theta + x/\overline{d})>\, .
\ee
The average is taken over the unit circle, i.e.\ 
$<.> = \frac{1}{2\pi}\int_0^{2\pi} .\; d\theta \;$ which leads to
\be{two-point2}
R_2(x,N) = \frac{1}{N} \sum_{i,j=1}^N \dpi(x_i - x_j - x) 
= 1 + \frac{2}{N^2} \sum_{n=1}^{\infty} |\Tr \vec{U}^n(N)|^2 
\cos(2\pi \frac{n}{N} x) \, ;
\ee
the spectrum $\{x_i = \overline{d}\, \theta_i; \, i=1,2\ldots N\}$ is unfolded
to mean level density one. Note that $R_2$, as defined in (\ref{two-point1}), 
is still a distribution and further averaging has to be applied. Fourier 
transformation of the 
two-point correlation function yields the so-called form factor which 
can be written as
\be{K1}
K(\tau,N) = \left\{ \begin{array}{lll}
\frac{1}{N} |\Tr \vec{U}^{N\tau}(N)|^2  
\qquad &\mbox{for}& \quad \tau \ne 0\\
N  \qquad &\mbox{for}& \quad \tau = 0\, ;
\end{array} \right.
\ee
the variable $\tau$ takes the discrete values $\tau = n/N$, $n$, $N \in \N$ 
which reflects the periodicity of the two-point correlation function 
$R_2$. Note that the form factor does not converge in the limit $N\to\infty$ 
but fluctuates around a mean value. The limit distribution of $K(\tau,N)$ for
$N \to \infty$ is, however, expected to exist and is conjectured to be Gaussian
(Prange 1997).  I will refer to the form factor as the mean value of the 
distribution (\ref{K1}) near a given $\tau$ value, i.e.\ further averaging 
over small $\tau$-intervals has to be performed. 

The traces of $\vec{U}^n$ can in semiclassical approximation be written as 
sum over periodic orbits of topological length $n$ of the underlying 
classical map (Gutzwiller 1990, Bogomolny 1992), i.e.\
\be{Tr-sc}
\Tr \vec{U}^n(N) \approx \Tr \vec{\tU}^n(N) 
:= \sum_{p}^{(n)} A_{p} e^{2 \pi i N S_p} \, .
\ee
The complex prefactors $A_p$ contain information about the stability 
of the periodic orbit and $S_p$ denotes the action or generating function
of the map corresponding to the periodic orbit $p$. 
The relation between quantum traces and semiclassical periodic orbit sums is
in general obtained by stationary phase approximation and is exact 
only in the limit $\tau^{-1} = N/n \to \infty$ (Andersson and Melrose 1997).
Exceptions thereof are e.g.\ the cat map (Hannay and Berry 1980, Keating 1991) 
or quantum graphs (Kottos and Smilansky 1997) for which Eq.~(\ref{Tr-sc}) is 
exact for all $\tau$. 

The relation (\ref{Tr-sc}) makes it possible to connect purely classical action 
correlation functions with the statistics of quantum eigenvalues. A weighted
action correlation function can be obtained by considering 
\be{act-corr}
P(s,n) = \sum_{N = -\infty}^{\infty} |\Tr \vec{\tU}^n(N)|^2 e^{2\pi i s N} = 
\sum_{p,p'}^{(n)} A_p A_{p'}^* \, \dpi(S_p - S_{p'} - s) \; 
\ee 
which is a Fourier sum in $N$ depending only on classical quantities such 
as the topological length of the orbit, $n$, and the actions $S_p$ and weights
$A_p$.
Eqs.~(\ref{two-point2}) and (\ref{act-corr}) together with (\ref{K1}) 
indicate that energy level statistics and periodic orbit action 
correlations are connected provided that both the quantum traces 
$\Tr \vec{U}^n$ and the semiclassical 'traces' $\Tr \vec{\tU}^n$  follow 
the same probability distributions, i.e.\ random 
matrix statistics. Little is known about the statistical properties of 
semiclassical traces and eigenvalues and I will study the existence of 
universal RMT-behavior for semiclassical expressions in more detail in
section \ref{sec:sec4}. \\

The RMT-conjecture implies for quantum traces 
$\Tr \vec{U}^n(N)$ of systems whose classical limit is chaotic and non-time 
reversal symmetric 
\be{Krmt}
\lim_{N\to\infty}\frac{1}{N}<|\Tr \vec{U}^n(N)|^2> = 
\left\{ \begin{array}{lll}
n/N   \qquad &\mbox{for}& \quad  n \le N\\
1  \qquad &\mbox{for}& \quad n > N
\end{array} \right.  ;
\ee
the brackets $<.>$ indicate the average over an $n$-interval small compared 
to the dimension $N$. 
Assuming that the above relation holds also for the semiclassical traces  
$<|\Tr \vec{\tU}^n(N)|^2>$
we can insert (\ref{Krmt}) in (\ref{act-corr}) which leads to 
the asymptotic result (Argaman \et 1993)
\be{Prmt1}
P(s,n) = \overline{P}(n) 
- \left( \frac{\sin n \pi s}{\pi s} \right)^2 + n \delta(s) 
\qquad \mbox{for} \quad |s| \ll 1 .
\ee

Here, $\overline{P}(n) = \sum_{p,p'}^{(n)} A_p A_{p'}^*$ is the mean part of
the weighted periodic orbit action pair density.
Note that the number of periodic orbits increase exponentially with
$n$ for chaotic maps which in turn implies an exponential increase for the 
density of periodic orbit actions (modulus 1) and in general also for 
the weighted density $\overline{P}(n)$.

After rescaling the $s$ - variable according  to $s = \sigma/n$, 
Eq.~(\ref{Prmt1}) can be written as
\be{Prmt2}
P^{\small GUE}_{\small scal} (\sigma) = \frac{1}{n^2} 
P\left(\frac{\sigma}{n}\right) = \frac{\overline{P}(n)}{n^2} - 
\left(\frac{\sin \pi \sigma}{\pi \sigma} \right)^2 + \delta(\sigma) 
\qquad \mbox{for} \quad |\sigma| \ll n ,
\ee
which puts the non-trivial, $\sigma$-dependent correlations into $n$ - 
independent form.

The corresponding equation for systems with time-reversal symmetry is 
(Argaman \et 1993)
\begin{eqnarray} \label{Prmt3}
P^{\small GOE}_{\small scal} (\sigma) =
\frac{\overline{P}(n)}{n^2} &-& 
   2 \left(\frac{\sin \pi \sigma}{\pi \sigma} \right)^2 \nonumber\\
 &+&\frac{2}{\pi \sigma}\left[\cos 2\pi \sigma\left(\mbox{si}(2 \pi \sigma) 
\cos 2\pi \sigma 
  -  \mbox{Ci}(2 \pi \sigma) \sin 2\pi \sigma\right) 
   \right.\nonumber \\
  &+& \left. \mbox{Ci}(4 \pi \sigma) \sin 4\pi \sigma - 
       \mbox{si}(4 \pi \sigma) \cos 4\pi \sigma \right] + 2 \delta(\sigma)
\qquad \mbox{for} \quad |\sigma| \ll n ,
\end{eqnarray} 
with si($x$), Ci($x$) being the Sine - and Cosine - integrals.

For integrable systems conjectured to have Poissonian spectral statistics one
expects 
\[
\frac{1}{N}<|\Tr \vec{U}^n(N)|^2> = 
<\sum_{p,p'}^{(n)} A_p A_{p'}^* e^{2\pi i N (S_p - S_p')}> = 1 
\qquad \mbox{for} \quad |N| \ne 0\]
and thus 
\be{Ppois}
P^{\small Pois} (s,n) 
 = \sum_{N = -\infty}^{\infty} \frac{1}{N}
|\Tr \vec{\tU}^n(N)|^2 e^{2\pi i s N} 
= \overline{P}(n) - 1 + \delta(s),
\ee
i.e.\ periodic orbit actions are uncorrelated.

The $\delta$-functions in (\ref{Prmt1}) -- (\ref{Ppois}) can be identified with
the diagonal terms $p = p'$ in the sum (\ref{act-corr}) after making use of 
the classical sum rule (Hannay and Ozorio de Almeida 1984) 
\be{sum-rule} \sum_p^{(n)} |A_p|^2 \to  1 \qquad \mbox{for large } n \, .\ee

Before investigating the existence of correlations of the form (\ref{Prmt2})
or (\ref{Prmt3}) in more detail, I would like to make a few remarks: 
the weighted periodic orbit correlation function (\ref{act-corr}) contains
a background term $\overline{P}(n)$ which increases in general exponentially 
with $n$. This is in contrast to the linear behaviour of the mean quantum level
density $\overline{d}(N)$ in (\ref{density}) and (\ref{two-point1}).
Note also, that the complex weights $A_p$ carry phases which may 
result in cancellations in $\overline{P}$ and may occasionally lead to 
$\overline{P}$ = 0. 
The correlation functions (\ref{Prmt2}) and (\ref{Prmt3}) are in addition
\underline{not} rescaled with respect to the mean action density as e.g the 
two-point spectral correlation function $R_2$ in (\ref{two-point1}). 
The differences in action between adjacent orbits of the same length $n$ is 
exponentially small on the $s$ or $\sigma$ scale and the term 
$ - (\sin \pi \sigma/\pi \sigma)^2$  in (\ref{Prmt2}), 
(\ref{Prmt3}) indicates long range correlations on the scale of the 
mean periodic action separation. 
The correlations (\ref{Prmt2}) and (\ref{Prmt3}) imposed by random matrix 
theory are thus a small modulation of order ${\cal O}(1)$ on top of an 
exponentially large background when considering all periodic orbit pairs 
(Dahlqvist 1995).  The two-point correlation function for periodic orbit 
actions approaches the Poissonian limit for $n\to\infty$ when rescaling 
$P(\sigma,n)$ with respect to the 
mean action density in accordance to e.g.\ the definition of the spectral 
correlation function (\ref{two-point1}); periodic orbits are thus 
uncorrelated on scales of the mean action separation in agreement with the 
numerical results by Harayama and Shudo (1992). 

The search for an  ${\cal O}(1)$ -- effect on top of an exponentially large 
background is one of the main obstacles when investigating RMT-induced 
classical action correlations numerically. The second major problem in 
studying periodic orbit properties especially in the large $n$ - limit is 
the exponential 
increase in the number of orbits with increasing $n$. In the following I will 
focus on a specific example, the classical and quantum Baker map,
in which the problems mentioned earlier can be overcome by constructing a 
quasiclassical Perron--Frobenius type operator (Dittes \et 1994). 

\section{The Baker Map}
\label{sec:sec3}
The Baker map has become a standard example when studying classical and quantum 
chaos (see e.g.\ Balazs and Voros 1989, Saraceno and Voros 1994, 
O'Connor \et 1992, Hannay \et 1994). The classical dynamics 
is given by a two-dimensional, piecewise linear, area-preserving map on the 
unit square defined as
\begin{eqnarray}\label{b-map}
q'& =& 2 q - \epsilon \nonumber \\
p' &=& \frac{1}{2} (p + \epsilon) \qquad \mbox{with} \quad \epsilon =[2q] 
\end{eqnarray}
and $(q,p)$, $(q',p')$ being the initial and final points. The notation $[x]$ 
stands for the integer part of $x$. The Baker map is hyperbolic with a well 
defined Markov - partition and a complete binary symbolic dynamics given in 
terms of the $\epsilon = \{0,1\}$. The map is invariant under the 
anti-unitary symmetry $T:(q,p) \to (p,q)$ (being equivalent to time reversal 
symmetry) and a parity transformation $R:(q,p) \to (1-q,1-p)$. 
A proper desymmetrisation of quantum spectra and semiclassical periodic orbit 
sums with respect to parity will be crucial when studying 
spectral statistics and periodic orbit correlations in section 
\ref{sec:sec4}. 

Each periodic orbit of the map can be associated with a finite symbol string
$\vecgr{\epsilon} = (\epsilon_1, \epsilon_2, \ldots, \epsilon_n)$ and the phase
space coordinates of the orbit are given by the expression
\be{po-coord}
q_{\vecgr{\epsilon}} = \frac{1}{2^n - 1} \sum_{i=0}^{n-1} 
\epsilon_i 2^{n-i-1}; \quad
p_{\vecgr{\epsilon}} = \frac{1}{2^n - 1} \sum_{i=0}^{n-1} \epsilon_i 2^{i}\, .
\ee

A suitable generating function of the map is 
\be{gen-fun}
 W_{\epsilon} (p',q) = 2p'q - \epsilon p' - \epsilon q 
\qquad \mbox{with} \quad \epsilon =[2q]\, , 
\ee 
which in turn defines the action; the action of a periodic orbit 
${\vecgr{\epsilon}}$ of the map can up to an additive integer be written 
as (Dittes \et 1994)
\be{po-action}
S_{\vecgr{\epsilon}} = \sum_{i=1}^n (q_i-1)\epsilon_i = 
\sum_{i=1}^n (q_i-1)[2 q_i] 
\ee
with $q_i$ being the $q$-coordinates of the periodic orbit. The sum in 
(\ref{po-action}) is invariant under time reversal symmetry 
and parity-transformation $R(q_i) = 1-q_i$; the later can be shown using the 
relation $\sum_{i=1}^n q_i = \sum_{i=1}^n \epsilon_i$. 

A quantized version of the Baker map is obtained by making use of 
a discretised version of the generating function (\ref{gen-fun}) and the
classical structure of the map in the mixed representation. A suitable 
formulation preserving all the classical symmetries is provided by the
choice (Balazs and Voros 1989, Saraceno and Voros 1994)
\be{b-map-qm}
\vec{U}(N) = \vec{F}^{-1}_{N} \times \left( \begin{array}{cc}
				    \vec{F}_{N/2}&   0 \\
	                              0    &  \vec{F}_{N/2}
				   \end{array} \right)
\ee
with 
\be{pq-trafo}
(\vec{F}_N)_{\rm ij} = \frac{1}{\sqrt{N}} 
                     e^{-2 \pi i ({\rm i}-\frac{1}{2})({\rm j}-\frac{1}{2})/N}
\qquad {\rm i,j} = 1, \ldots N \; .
\ee
The Fourier-matrix $\vec{F}$ is nothing but the transformation from position 
to momentum representation in the finite dimensional Hilbert space. The 
dimension $N$ of the map $\vec{U}(N)$ has to be even in this construction 
and $N$ is equivalent to the inverse of Planck's constant. The unitary map 
$\vec{U}(N)$ commutes with the parity operator $(\vec{R}_{N})_{\rm i,j} = 
\delta_{{\rm i},N-{\rm j}+1}$ due to the particular choice of half-integer 
phases 
(Saraceno and Voros 1994). Symmetry reduction is obtained by considering 
the matrices $\vec{U}_{\pm}(N)$ defined as
\be {b-map-pm}
\vec{U}_{\pm}(N) = \frac{1}{2}
\left(\vec{U}(N) \pm \vec{U}(N) \vec{R}_{N}\right)
\ee
acting on the symmetric/anti-symmetric wave-vectors only. This leads 
effectively to a reduction in dimension by a factor of 2. The traces of 
$\vec{U}_{\pm}^n$ can in the large $N$-limit be written as 
(Saraceno and Voros 1994, Toscano \et 1997)
\be{Tr-b-N}
\Tr\vec{U}_{\pm}^n(N) = \frac{1}{2} \sum_{\vecgr{\epsilon}}^{(n)}\left( 
\frac{2^{n/2}}{2^n-1} e^{2\pi i N S_{\vecgr{\epsilon}}} 
\pm \frac{2^{n/2}}{2^n+1} e^{2\pi i N S'_{\vecgr{\epsilon}} + i \pi} + 
a_{\vecgr{\epsilon}} \log N + b_{\vecgr{\epsilon}}\right) + 
{\cal O}(N^{-1/2}) + \ldots \, .
\ee
The first two terms in the sum over the possible finite symbol strings of
length $n$ are the usual semiclassical Gutzwiller-like periodic orbit 
contributions. They arise as stationary phase points in a continuum 
approximation of $\Tr \vec{U}^n$, $\Tr(\vec{U}^n \vec{R})$, 
respectively, i.e.\ sums over matrix elements are replaced by integrals 
(Tabor 1983). The action $S_{\vecgr{\epsilon}}$ of a 
periodic orbit of length $n$ is given by Eq.\ (\ref{po-action}). 
$S'_{\vecgr{\epsilon}}$ corresponds to half the action of 
an orbit of length $2n$ with symbol code ($\epsilon_1, \ldots \epsilon_n, 
\bar{\epsilon_1}, \ldots \bar{\epsilon_n}$) and $\bar{\epsilon_i} = 
1 - \epsilon_i$. These are the orbits being invariant under the classical 
parity transformation $R$ and are hence the stationary phase contributions of 
$\Tr (\vec{U}^n \vec{R})$.  The anomalous 
$a_{\vecgr{\epsilon}} \log N + b_{\vecgr{\epsilon}}$
corrections are due to the discretisation of the phase-space and arise
in 'sub leading' integrals when performing Poisson-summation on $\Tr\vec{U}^n$ 
i.e.\ replacing sums by sums over integrals (Saraceno and Voros 1994,
Toscano \et 1997)). Next leading terms are diffraction corrections 
arising from the discontinuous nature of the classical and quantum map
(\ref{b-map}) and (\ref{b-map-qm}). Note the extra phase $\pi$ in the 
$\Tr(\vec{U}^n \vec{R})$ - contributions which originates from the discretised
stationary phase approximation. 

In the following I will mainly concentrate on the Gutzwiller - like
periodic orbit contributions, i.e.\ I will consider the approximation 
\begin{eqnarray}\label{Tr-b-sc}
\Tr \vec{U}^n(N) &\approx& \Tr \vec{\tU}^n(N) 
= \frac{2^{n/2}}{2^n-1} \sum_{\vecgr{\epsilon}}^{(n)} 
e^{2 \pi i N S_{\vecgr{\epsilon}}}\\
\Tr \vec{U}_{\pm}^n(N) &\approx& \Tr \vec{\tU}_{\pm}^n(N) 
= \frac{1}{2}\left[
\frac{2^{n/2}}{2^n-1} \sum_{\vecgr{\epsilon}}^{(n)} 
e^{2 \pi i N S_{\vecgr{\epsilon}}} \mp
\frac{2^{n/2}}{2^n+1} \sum_{\vecgr{\epsilon}}^{(n)} 
e^{2 \pi i N S'_{\vecgr{\epsilon}}} \right] \; .
\nonumber
\end{eqnarray}
I will henceforth refer to the periodic orbit sums (\ref{Tr-b-sc}) as
the semiclassical approximation of the quantum Baker map.
The influence of the remaining contributions in (\ref{Tr-b-N}), 
including the leading order ($\log N$) - corrections, will be discussed at 
the end of this section.\\

The main problem in calculating periodic orbit sums like (\ref{Tr-b-sc}) is 
the exponential increase of periodic orbit contributions. 
Even for the Baker map where periodic orbit actions are given 
by the simple analytic formula (\ref{po-action}) 
a direct summation of $\Tr \vec{\tU}^n$ 
becomes a computational challenge for $n\ge30$. 
For the Baker map it is, however, possible, to construct a quasiclassical 
operator $\tU$ whose traces coincide with the semiclassical periodic orbit 
sum (\ref{Tr-b-sc}) (Dittes \et 1994). The spectrum of this operator 
can be computed explicitly and simplifies the 
task of calculating traces, i.e.\ periodic orbit sums, considerably.
Other quasiclassical operators for the Baker map have been proposed by 
Kaplan and Heller (1996), Sano (1999) and Hannay (1999). Note, that an 
identification of semiclassical periodic orbit sums with classical or 
quasiclassical operator is not possible for generic systems and 
the notation $\Tr \vec{\tU}^n$ is then a mere substitution 
for the periodic orbit sum itself. 

The Baker map is special for the following reasons: firstly, the periodic 
orbit amplitudes $A_p$, (cf.\ Eq.~(\ref{Tr-sc})), do not depend on the 
specific orbit, but only on the topological length $n$, which is a 
consequence of the piecewise linearity of the classical map. Secondly, the 
dynamics of the $q$ - coordinate as well as the periodic orbit - action 
(\ref{po-action}) are independent of the momentum $p$ which allow for a 
classical separation 
of momentum and configuration space variables. As a consequence, one can define 
a one dimensional Perron-Frobenius type integral kernel acting on the 
$q$-coordinate only (Dittes \et 1994). In a form preserving the 
parity-symmetry $R(q) = 1-q$ it can be written as
\be{U-qcl}
\tU(q,q';N) = \sqrt{2}\; \delta\left(q - (2q'-[2q'])\right) 
e^{2 \pi i N S(q')} 
\ee
with 
\be{S-qcl}
S(q') = q'[2q']-\frac{1}{2}([2q'] + q') \; .
\ee
One easily deduces $\tU(q,q') = \tU(Rq,Rq')$. The traces of $\tU(q,q')$
coincide with the periodic orbit sum (\ref{Tr-b-sc}) if the operator is
defined on the space of analytic functions on the unit interval with 
periodic boundary conditions (Dittes \et 1994, Rugh 1992). The operator
is thus infinite dimensional (having infinitely many eigenvalues) in contrast
to the finite dimensional quantum matrix $\vec{U}(N)$ and $N$ is a mere 
parameter in (\ref{U-qcl}). A symmetry-reduced version of this quasiclassical
operator is obtained by considering
\be{U-qcl-pm}
\tU_{\pm}(q,q';N) = \frac{1}{2}\left(\tU(q,q';N) \mp \tU(q,Rq';N)\right)\; .
\ee
Note, that the quasiclassical operator acting on antisymmetric functions 
(denoted $\tU_{+}$ here) corresponds to the symmetric quantum subspace 
$\vec{U}_{+}$ and vice versa, reflecting the `quantum' origin of the extra phase
for the periodic orbit contributions to $\Tr \vec{U}^n\vec{R}$, see 
Eq.~(\ref{Tr-b-N}). The operator (\ref{U-qcl}) or (\ref{U-qcl-pm}) can now
be represented as an infinite dimensional matrix after choosing a suitable 
basis set and the calculation of traces can be reduces to matrix calculus
(provided the operator is trace class). An obvious choice for the basis 
functions is the Fourier-basis, i.e.\
\begin{eqnarray}\label{U-qcl-matrix}
\tU_{k,m} &=& \int_{0}^{1}dq \int_{0}^{1}dq' \tU(q,q';N) e^{2\pi i (mq'-kq)}\\
          &=& (-1)^m \sqrt{2} \frac{e^{-\pi i N/2}}{2 \pi i} 
	      \left[\frac{e^{\pi i (N/2 + m - 2 k)} - 1}{N/2 + m - 2 k} + 
	      \frac{e^{\pi i (N/2 - m + 2 k)} - 1}{N/2 - m + 2 k}\right] \; .
\nonumber
\end{eqnarray}
Similarly one obtains
\begin{eqnarray}\label{U-qcl-matrixI}
(\tilde{U}R)_{k,m} &=& 
\int_{0}^{1}dq \int_{0}^{1}dq' \tU(q,Rq';N) e^{2\pi i (mq'-kq)}\\
          &=& (-1)^m \sqrt{2} \frac{e^{-\pi i N/2}}{2 \pi i} 
	      \left[\frac{e^{\pi i (N/2 + m + 2 k)} - 1}{N/2 + m + 2 k} + 
	      \frac{e^{\pi i (N/2 - m - 2 k)} - 1}{N/2 - m - 2 k}\right] \;
\nonumber
\end{eqnarray}
with $k,m$ integers. The various terms 
$(e^{\pi i (N/2 \pm m \pm 2 k)} - 1)/(N/2 \pm m \pm 2 k)$ are equal to 
$i \pi$ for $N/2 \pm m \pm 2 k = 0$.

%%%%%%%%%%%%%%%%%%%%%%%%%%%%%%%%%%%%%%%%%%%%%%%%%%%%%%%%%%%%%%%%%%%%%%%%%%%
\begin{figure}
\centering
\centerline{
         \epsfxsize=15.0cm
         \epsfbox{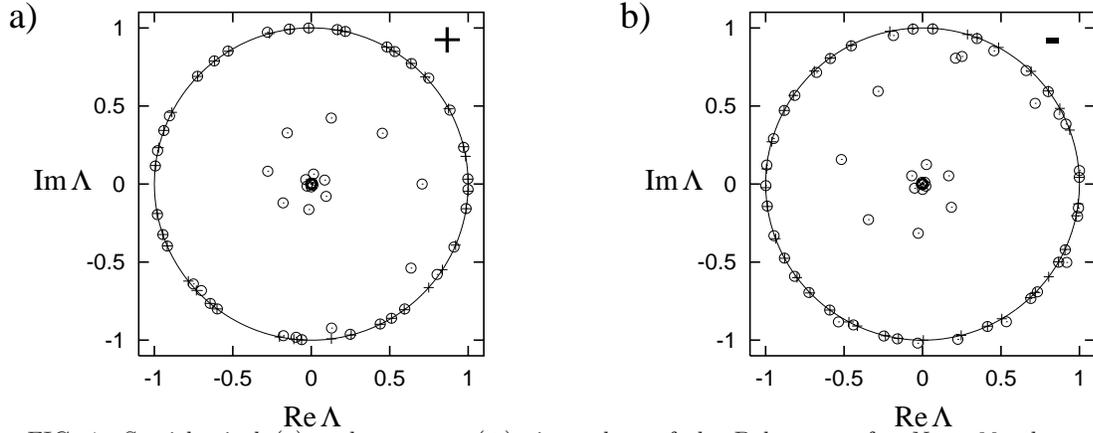}
         }
\caption[]{\small
Semiclassical ($\circ$) and quantum (+) eigenvalues of the Baker map for
$N = 80$; the parity subspaces are shown separately: a) positive parity, 
b) negative parity.
}
\label{Fig:spec}
\end{figure}
%%%%%%%%%%%%%%%%%%%%%%%%%%%%%%%%%%%%%%%%%%%%%%%%%%%%%%%%%%%%%%%%%%%%%%%%%%%

A typical eigenvalue spectrum $\{\Lambda_i\}, i=0,1,2 \ldots$ of the 
quasiclassical operator $\tU_{\pm}$ is shown in Fig.~\ref{Fig:spec} together 
with the quantum eigenvalues. In the following, the eigenvalues will be ordered
with decreasing modulus, i.e.\ $|\Lambda_i|>|\Lambda_j|$ for  
$i < j$. The operator $\tU$ approximates unitarity of the 
corresponding quantum map $\vec{U}$ quite well; approximately $N/2$ 
eigenvalues in each parity subspace lie near the unit circle and 
these eigenvalues agree well with quantum results. The modulus of semiclassical 
eigenvalues becomes exponentially small for $i > N/2$ in each subspace and
infinitely many eigenvalues `disappear' by spiraling into the origin. 
This behavior 
reflects the trace-class property of 
the operator and makes it possible to truncate the size of the matrix 
$\vec{\tU}_{\pm}$ in actual calculations. (Choosing dim$(\vec{\tU}_{\pm})=
3\times\frac{N}{2}$ ensures convergence of the first $N/2$ eigenvalues in 
each parity subspace to 4 significant digits). \\

%%%%%%%%%%%%%%%%%%%%%%%%%%%%%%%%%%%%%%%%%%%%%%%%%%%%%%%%%%%%%%%%%%%%%%%%%%%
\begin{figure}
\centering
\centerline{
         \epsfxsize=8.0cm
         \epsfbox{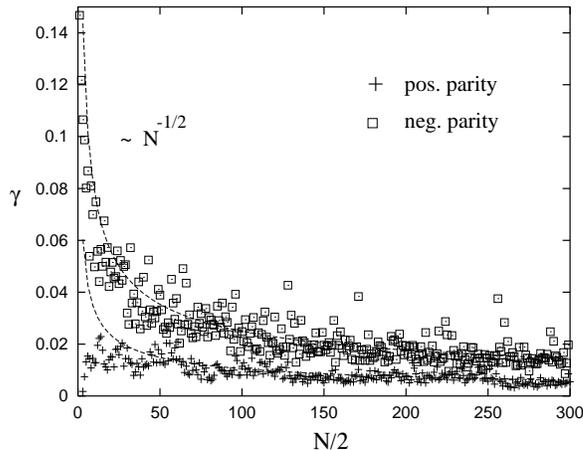}
         }
\caption[]{\small
The semiclassical eigenvalue with the largest modulus plotted here as 
$\gamma^{\pm}(N) = \log |\Lambda^{\pm}_0(N)|$ versus $N/2$ for
both subspaces; the deviation from the unit circle decreases in both
cases like $1/\sqrt{N}$ (dashed lines), but one finds typically 
$\gamma^{+}(N)< \gamma^{-}(N)$. 
}
\label{Fig:gmax}
\end{figure}
%%%%%%%%%%%%%%%%%%%%%%%%%%%%%%%%%%%%%%%%%%%%%%%%%%%%%%%%%%%%%%%%%%%%%%%%%%%

A measure for the error of the semiclassical approximation compared to 
quantum calculations is the deviation of the semiclassical eigenvalue with 
largest modulus, $\Lambda_{0}(N)$, from the unit circle. The distribution of 
$\gamma^{\pm}(N) := \log |\Lambda^{\pm}_0(N)|$
for integer values of $N$ and both subspaces is shown in Fig.~\ref{Fig:gmax}. 
The deviation of semiclassical eigenvalues from the unit - circle decreases 
with increasing $N$ following a $1/\sqrt{N}$ behavior. The $1/\sqrt{N}$ - 
scaling suggest that corrections to individual semiclassical eigenvalues are 
dominated by diffraction contributions. The $\log(N)$ corrections in the 
traces, cf.~(\ref{Tr-b-N}), become dominant when summing over the periodic 
orbit contributions or equivalently over the semiclassical eigenvalues 
and are thus a collective effect of the various contributions.

The deviations from the unit circle for semiclassical eigenvalues
with modulus greater than one is systematically larger for eigenvalues 
corresponding to the $\tU_-$ - operator compared to those from the 
$\tU_+$ - operator. This can qualitatively be understood for small $N$ when 
considering the limit $N = 0$; the operator $\tU(q,q';N=0) = \sqrt{2}\, 
\delta\left(q - (2q'-[2q'])\right)$ is the classical Perron-Frobenius 
operator for the Sawtooth-map up to a factor of $\sqrt{2}$. The spectrum 
$\tU(q,q';0)$ has therefor a single non-zero eigenvalue $\sqrt{2}$ 
in the symmetric subspace which corresponds to $U_-$ in our notation 
all other eigenvalues are zero. The eigenvalues are smooth functions 
of the continuous variable $N$ which leads to $|\Lambda_{0}^{-}| > 
|\Lambda_{0}^{+}|$ for $N$ smaller one. The fact that this behavior persists 
for large $N$ values , cf.\ Fig.\ \ref{Fig:spec}, can certainly not be 
explained by the argument above but is worth noting as a purely numeric 
result.

\section{Semiclassical spectral statistics and action correlation functions:
numerical results}
\label{sec:sec4}
The existence of a quasiclassical operator for the Baker map makes it 
possible to write traces as convergent sums over the
quasiclassical eigenvalue spectrum, i.e.\
\be{tr-b-ev}
\Tr\tU_{\pm}^n(N) = \sum_{i=0}^{\infty} \left(\Lambda^{\pm}_i(N)\right)^n \, .
\ee
The violation of unitarity and the existence of semiclassical eigenvalues
with modulus greater than one leads to the asymptotic result
\[|\Tr\tU^n_{\pm}(N)| \approx  \exp(n \gamma^{\pm}(N)) \qquad \mbox{for}
\quad n \to \infty \]
with $\gamma^{\pm}(N) = \log |\Lambda^{\pm}_0(N)|$,
i.e.\ the traces grow exponentially in the limit $n\to\infty$ (Keating 1994,
Aurich and Sieber 1994). This kind of behavior is connected to the notorious 
convergence problems for periodic orbit sums of the form 
\[\sum_{n=1}^{\infty} \sum_{p}^{(n)} A_p e^{2 \pi i N S_p}\]
which are absolutely convergent only for complex $N$ with $Im(N)> h_t/2$ 
and $h_t$ is the topological entropy (Eckhard and Aurell 1989). For the Baker 
map the absolute sum over periodic orbit contributions (\ref{Tr-sc})
corresponds to $\sum_{n=1}^{\infty} \Tr\tU^n(0)$; this sum diverges like 
$2^{n/2}$ for large $n$, i.e.\ $h_t = \log 2$ here. 
A summation over the semiclassical eigenvalues (\ref{tr-b-ev}) gives a 
detailed account of the regime of conditional convergence for periodic orbit 
sums as a function of the real part of $N$. 

The exponential increase in the traces (\ref{tr-b-ev}) leads to
exponentially increasing terms in a semiclassical approximation of 
the form factor
\be{K-b-ev}
K_{sc}(\tau,N) = \frac{1}{N} |\Tr\tU_{\pm}^{N\tau}|^2 = 
\frac{1}{N}\sum_{i,j=0}^{\infty} 
\left(\Lambda^{\pm}_i \Lambda_j^{{\pm}^*}\right)^{N\tau} 
\approx \exp(2 N \tau \gamma^{\pm}(N)) \qquad \mbox{for} 
\quad \tau\to\infty \, .
\ee
A stationary distribution of $K_{sc}$ for fixed $\tau$ and $N\to\infty$ is 
obtained if $<\gamma(N)>$ falls off faster than  $N^{-1}$. Such a behavior is 
expected for smooth hyperbolic maps where the semiclassical error is dominated 
by $1/N$ - corrections due to higher order terms in a stationary phase 
expansion.  This is, however, not the case for the Baker map where diffraction 
effects of the order $N^{-1/2}$ dominate, see Fig.\ \ref{Fig:gmax}. 

Using the semiclassical traces $\Tr\tU^n$ to evaluate the two-point 
correlation function (\ref{two-point2}) will, on the other hand, always 
be affected by the exponentially increasing terms in the form factor; 
a semiclassical calculation of $R_2(x,N)$ diverges thus independently of 
$N$ and the behavior of $<\gamma(N)>$.\\

The weighted classical correlation function (\ref{act-corr}) is a finite
sum and thus well defined for fixed topological length $n$. From (\ref{Tr-b-sc})
one obtains in the large $n$ - limit
\begin{eqnarray*} 
\overline{P}(n) &=& 2^{-n} \qquad \; \mbox{in the `$+$' - subspace}\\
\overline{P}(n) &=&  2^{n}\qquad \quad \mbox{in the `$-$' - subspace}
\end{eqnarray*} 
and we will discard this 'trivial' mean part from now on. The large 
$n$ - asymptotics of the non-trivial correlations in (\ref{act-corr})
is dominated by the eigenvalue being the largest among all $\Lambda_0(N)$; 
after defining 
\[\gamma^{\pm}_{0} := \log|\Lambda_0^{\pm}(N_0)| := \sup_{N\in\sN} 
(\log|\Lambda_0^{\pm}(N)|)\]
where $N_0$ is the $N$ - value corresponding to the largest eigenvalue 
$\Lambda_0$, one obtains
\be{P-limit}
P(s,n) \to e^{2 n \gamma^{\pm}_0} \cos(2\pi N_0 s) \to\infty \qquad \mbox{for} 
\quad n\to\infty \; .
\ee

It follows from the considerations above that a full description of 
quantum spectral statistics in terms of classical actions and amplitudes 
needs to incorporate unitarity into a semiclassical approximation 
(Keating 1994). I will come back to this point at the end of the section. 
Apart from that there seems to be little information in the asymptotic 
behavior of $K_{sc}(\tau,N)$ or $P(s,n)$. The rates $|\gamma^{\pm}(N)|>0$ 
correspond to `semiclassical escape rates' and contain little information 
about the classical dynamics or the quantum map. The distribution of these 
rates is instead of statistical nature due to accumulation of errors in the 
semiclassical approximation. 

\begin{figure}
\centering
\centerline{
         \epsfxsize=15.0cm
         \epsfbox{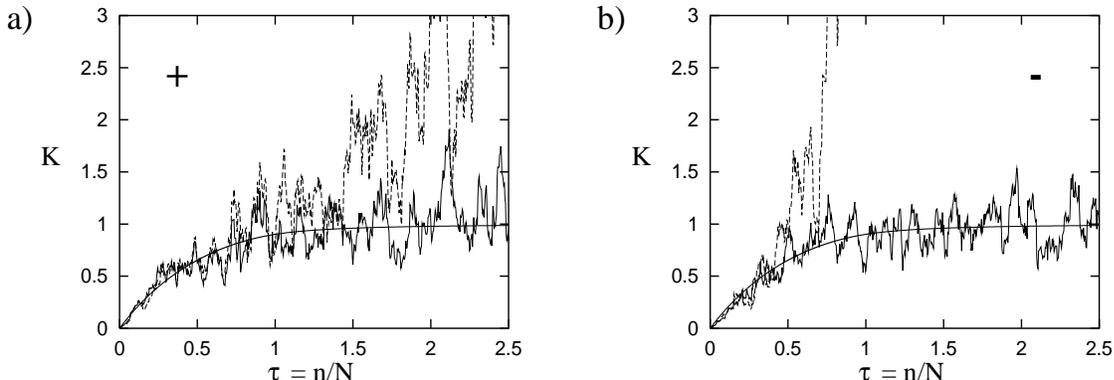}
         }
\caption[]{\small
The form factor $K(\tau)$, i.e.\  the square modulus of the quantum traces
$\Tr \vec{U}^n$ (full line) and the semiclassical traces $\Tr \tU^n$ (dashed line)
for positive (a) and negative (b) parity is displayed as function of 
$\tau = n/N$ for $N/2 = 259$ together with the RMT - result for the 
Gaussian Orthogonal Ensemble (smooth curve). The semiclassical break-times,
Eq.~(\ref{break}), are $\tau^+_b \approx 1.002$, $\tau^-_b \approx .301$,
respectively. 
}
\label{Fig:k-all}
\end{figure}

The more interesting question in this context is whether or not semiclassical 
periodic orbit formulae for classical chaotic systems are able to reproduce 
RMT-statistics quantitatively in the non-asymptotic regime. The form factor 
obtained from the quasiclassical operator is displayed in Fig.\ \ref{Fig:k-all}
together with the result obtained from the quantum spectrum and the 
GOE-prediction (applicable for systems with time reversal symmetry). 
The semiclassical result is indeed
capable of reproducing the RMT-result in both subspaces for small $\tau$ 
values; especially the deviation of the RMT-behavior from the classical
result $K(\tau) = 2 \tau$ obtained from the sum rule (\ref{sum-rule}) in the
limit $\tau \to 0$ is here well reproduced by periodic orbit formulae. 
This clearly indicates that there are non-trivial correlations between 
periodic orbit actions for systems with time-reversal symmetry. 

%%%%%%%%%%%%%%%%%%%%%%%%%%%%%%%%%%%%%%%%%%%%%%%%%%%%%%%%%%%%%%%%%%%%%%%%%%%
%%%%%%%%%%%%%%%%%%%%%%%%%%%%%%%%%%%%%%%%%%%%%%%%%%%%%%%%%%%%%%%%%%%%%%%%%%%

Representing
periodic orbit formulae by sums over quasiclassical eigenvalues allows one to 
study the large $N$ and $\tau$ limit
of a semiclassical approximation of the form factor. 
Exponentially increasing components start to dominate $K_{sc}(\tau, N)$ for 
$\tau$ larger than a semiclassical break-time of the order
\be{break}\tau^{\pm}_b \sim \frac{1}{\gamma^{\pm}(N) N} \ee
The break-time $\tau^{\pm}_b$ is typically a factor 5 larger in the
`$+$' subspace than in the `$-$' subspace, see Fig.~\ref{Fig:gmax}.\\

The action correlation function (\ref{act-corr}), on the other hand,
can be obtained either by sampling action-differences directly (which is
limited to $n<15$ -- 20 due to the exponential increase in the number of
orbits) or with the help of the traces $\Tr{\tU}^n(N)$. The latter method
demands calculation of the spectra of ${\tU}(N)$ for integer 
$N$-values and summation of the Fourier-sum in (\ref{act-corr}) directly. 
In practice,
the sum is truncated at a finite $N$-value, $N_{\smax}$, which 
corresponds to a smoothing of the original correlation function $P(s,n)$ 
on scales $1/N_{\smax}$; after rescaling according to Eq.~(\ref{Prmt2}) 
and subtracting $\overline{P}(n)$ one obtains
\be{conv}
P_{scal}(\sigma,n,N_{\smax}) = \frac{2}{n^2} \sum_{N=1}^{N_{\smax}} 
|\Tr \vec{\tU}^n(N)|^2 \cos(2\pi \frac{N}{n} \sigma) \approx 
\int_{-\infty}^{\infty} d\sigma' P_{scal}(\sigma',n) g_{\alpha}
(\sigma-\sigma')
\ee 
with 
\be{smooth}
g_{\alpha}(x) = \frac{\sin(2\pi \alpha x)}{\pi x} \qquad \mbox{and} \quad
\alpha = \frac{N_{\smax}}{n}.
\ee

\begin{figure}
\centering
\centerline{
         \epsfxsize=15.0cm
         \epsfbox{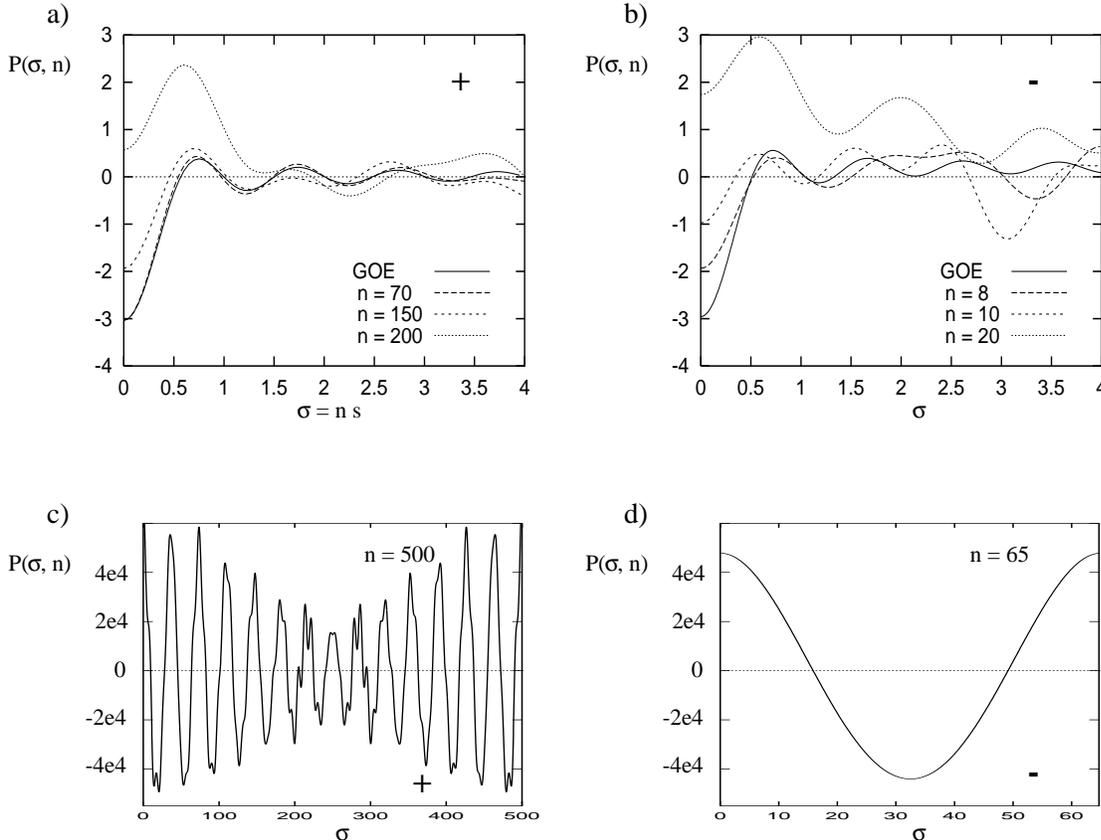}
         }
\caption[]{\small
The periodic orbit correlation function are shown for various $n$ -- values 
in the `$+$' subspace (Fig.\ a, c) and `$-$' subspace (Fig.\ b, d); 
'universal' 
periodic orbit correlation which coincide with the GOE - prediction can be
observed up to $n \approx 70$ in the  `$+$' subspace, but only up to 
$n \approx 8$ in the  `$-$' subspace. Non-universal periodic orbit
correlations develop for large $n$ (with exponentially increasing amplitude),
see Fig.\ c and  d. The cut-off is $N_{\smax} = n$ here, see 
Eq.~(\ref{conv}). 
}
\label{Fig:pcor}
\end{figure}

The smoothed periodic orbit correlation function $P_{scal}(\sigma,n,N_{\smax})$ 
is shown in Fig.~\ref{Fig:pcor} for various $n$-values and $N_{\smax} = n$.  
(Contributions from diagonal terms leading to the $\delta$ - functions in
(\ref{Prmt1}) and (\ref{Prmt2}) are subtracted here). 
Universal periodic orbit correlations in the `$+$' subspace is observed 
up to $n \approx 70$; these are $2^{70}$ distinct periodic orbits, a 
number inaccessible to pure periodic orbit calculations. One can study even 
longer orbits and finds exponentially 
increasing terms dominating the correlation function for $n$ values
above the transition point $n \approx 70$. The modulations in the 
periodic orbit pair density function $P(\sigma,n,n)$ are indeed orders of 
magnitudes larger than the RMT-correlations for the $2^{500}$ periodic orbits 
of length $n=500$, see Fig.\ \ref{Fig:pcor}c 

Things look similar in the `$-$' subspace, 
the deviation from universality occur, however, for much smaller 
$n$-values, i.e.\ at $n \approx 6$. The correlations which occur
for larger $n$ have the simple form $P(s,n) = e^{\gamma^-_0 n} 
\cos(2\pi s)$, see Fig.\ \ref{Fig:pcor}d for $n=65$. This oscillatory 
behavior has already been observed by 
Sano (1999) when studying periodic orbit correlations in the Baker map 
without separating the two symmetry subspaces. Sano's result can now be 
interpreted in terms of the spectrum of the quasiclassical operator 
(\ref{U-qcl}); the exponents $\gamma_0^{\pm}$ for the two different subspaces, 
which determine the asymptotic behavior of $P(s,n)$ for large $n$, 
see Eq.~(\ref{P-limit}), are
\begin{eqnarray*}
\gamma^+_0 = 0.022885 \qquad &\mbox{with}& \quad N_0 = 14\\
\gamma^-_0 = 0.148508 \qquad &\mbox{with}& \quad N_0 = 1 \, , 
\end{eqnarray*}
which can also be deduced from Fig.\ \ref{Fig:gmax}.
The exponentially growing terms in the periodic orbit correlation function 
have thus a seven times larger leading exponent in the `$-$' subspace compared 
to the `$+$' subspace. The correlation function $P(s,n)$ in the `$-$'  
subspace is dominated by the Fourier coefficient $|\Tr\tU^n(1)|^2$ 
for $n \ge 6$ and RMT-type correlation can not develop. The same is obviously 
true when considering the full Baker map without symmetry-reduction.\\

%%%%%%%%%%%%%%%%%%%%%%%%%%%%%%%%%%%%%%%%%%%%%%%%%%%%%%%%%%%%%%%%%%%%%%%%%%%
%%%%%%%%%%%%%%%%%%%%%%%%%%%%%%%%%%%%%%%%%%%%%%%%%%%%%%%%%%%%%%%%%%%%%%%%%%%

Universal correlations coinciding with the RMT-prediction do, however, exist.
This is clearly demonstrated in the `$+$' subspace for $n \le 70$. It is 
the violation of unitarity in a semiclassical approximation which
leads (in general) to exponentially diverging terms which in turn overwhelm
the pair-distribution function $P(s,n)$ for large $n$. The existence 
of universal periodic orbit correlations is thus strongly 
linked to the preservation of unitarity in a semiclassical formulation. 
The absence of RMT - like periodic orbit correlation, as observed in the 
`$-$' subspace, on the other hand, contains little information about the 
phenomena of 
universal spectral statistics but merely reflects the limitations of the 
approach due to the underlying semiclassical approximations. \\ 

A possibility to enforce unitarity onto a semiclassical
description has been proposed by Bogomolny and 
Keating (1996a). The starting point is the quantum staircase function 
${\cal N}(\theta) = \sum_{i=1}^N \Theta(\theta-\theta_i)$ with 
$\theta_i$ being the quantum eigenphases and $\Theta(x)$ denotes the 
Heaviside step-function. A smoothed, semiclassical version of the 
staircase function may be written as the truncated 
Fourier-series of ${\cal N}(\theta)$ 
\be{stair}
\tilde{\cal N}(\theta,N) = \frac{N}{2 \pi} \theta - 
\frac{1}{\pi}Im \sum_{n=1}^{N} 
\frac{1}{n}\Tr \vec{\tU}^n(N) e^{-i n\theta} \, ,
\ee
and the traces $\Tr \vec{\tU}^n$ are expressed as sums over periodic orbit 
contributions, see Eq.~(\ref{Tr-sc}). The unitarity condition 
is implemented by choosing a quantization condition
\be{quant}
\tilde{\cal N}(\tilde{\theta}_n,N) \stackrel{!}{=} n + \frac{1}{2} 
\ee
with $\tilde{\theta}_n$ being the solutions of (\ref{quant}); the 
$\tilde{\theta}_n$
represent a semiclassical approximation of the quantum eigenphases
$\theta_n$. A density of states is defined according to 
\be{density1}
D(\theta,N) =  \sum_{i=1}^N \dpi(\theta- \tilde{\theta}_i) 
            = \tilde{d}(\theta,N) \sum_{i=1}^N 
            \dpi\left(\tilde{\cal N}(\theta,N) - n - \frac{1}{2}\right) 
\ee
with $\tilde{d}(\theta,N) = \frac{\partial}{\partial \theta} 
\tilde{\cal N}(\theta,N)$. The density (\ref{density1}) is again 
written in terms of periodic orbit contributions only but  
`preserves' unitarity by construction. Bogomolny and Keating (1996a) 
started from the expression (\ref{density1}) to derive periodic orbit 
corrections to the quantum two-point correlation function $R_2$ beyond the 
results obtained from the classical sum-rule (\ref{sum-rule}).  

The Perron-Frobenius representation of the semiclassical traces for the 
Baker map allows one to test the quantization condition (\ref{quant}) for
large $N$ up to $N \approx 500$ by calculating the semiclassical traces 
in (\ref{stair}) directly; the `regularised' spectrum 
\{$\tilde{\theta}_i$\} 
thus obtained is used to construct new semiclassical traces 
$\Tr \tilde{ \tilde{U} }^n = \sum_i \exp(i n \tilde{\theta}_i)$ which
in turn are used to calculate the periodic orbit correlation function 
(\ref{act-corr}). Note, however, that the simple relation between periodic
orbits of length $n$ and the $n$-th semiclassical trace gets lost here and the
new periodic orbit correlation function $P(s,n)$ contains also correlation
between orbits and pseudo - orbits (being composites of shorter orbits).

\begin{figure}
\centering
\centerline{
         \epsfxsize=9.0cm
         \epsfbox{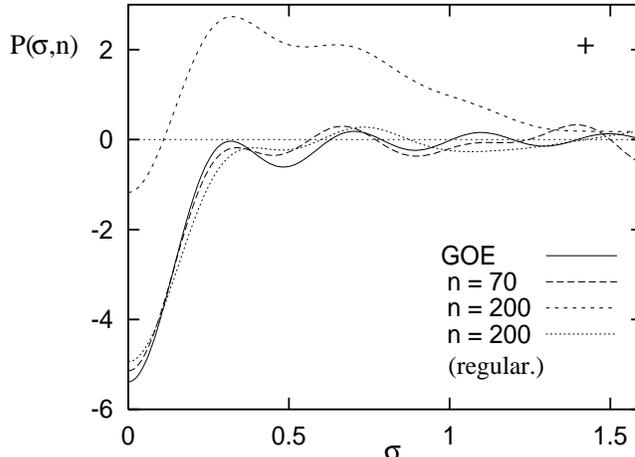}
         }
\caption[]{\small
The periodic orbit correlation function shown for various $n$ -- values 
in the `$+$' subspace with enhanced resolution compared to 
Fig.\ \ref{Fig:pcor}; the regularisation procedure (\ref{quant}) 
eliminates the exponentially growing terms and universal correlations
are uncovered now also for $n = 200$. 
}
\label{Fig:pcor-bk}
\end{figure}

The correlation function obtained from the `regularised' 
semiclassical data is shown in Fig.\ \ref{Fig:pcor-bk} together with
correlation functions where regularisation has not been applied, cf.\ also 
Fig.\ \ref{Fig:pcor}. The functions $P(\sigma,n,N_{\smax})$ are shown here 
with enhanced resolution 
compared to Fig.\ \ref{Fig:pcor}, i.e.\ with a cut-off $N_{\smax} = 
\frac{5}{2} n$. The periodic orbit correlation function for $n = 70$ follows 
the random matrix prediction also with enhanced resolution, but again deviates 
from the universal behavior for $n = 200$. 
The exponentially growing terms are eliminated in the corresponding regularised
correlation function using the spectrum obtained from (\ref{quant}) for 
$n = 200$. The regularisation procedure unveils the underlying universal 
correlations in the classical actions, see Fig.\  \ref{Fig:pcor-bk}.

The regularisation process (\ref{quant}) has the disadvantage that the 
Fourier-components of the new density of states 
$D(\theta,N$) can no longer be written as closed expressions of a finite 
number of periodic orbits; they become complicated infinite 
periodic orbit sums after writing the 
$\dpi$ -- function in its Fourier-components (Bogomolny and Keating 1996a). 
The resulting periodic orbit correlation functions, as e.g.\ shown in Fig.\ 
\ref{Fig:pcor-bk} for $n = 200$, consist of periodic orbit contributions 
of orbits and pseudo - orbits of topological length up to and including 
$n = 200$ which does not simplify the task of understanding 
the origin of universal periodic orbit correlations. 

%%%%%%%%%%%%%%%%%%%%%%%%%%%%%%%%%%%%%%%%%%%%%%%%%%%%%%%%%%%%%%%%%%%%%%%%%%%
%%%%%%%%%%%%%%%%%%%%%%%%%%%%%%%%%%%%%%%%%%%%%%%%%%%%%%%%%%%%%%%%%%%%%%%%%%%

\section{Conclusions and Outlook}
\label{sec:conl}
Correlations between actions of periodic orbits of the Baker map 
up to orbits of topological length $n=500$ (corresponding to 
$2^{500}$ different periodic orbits) have been studied with
the help of a quasiclassical Perron-Frobenius operator. 
The spectral form factor can be calculated by purely semiclassical 
expressions which coincide for small $\tau$ with random matrix results 
beyond the validity of the classical sum-rule, but diverge for 
$\tau \to \infty$. Action correlations of periodic orbits have been 
investigated 
which show universal non-trivial correlations linked to random matrix theory 
for short periodic orbits, but depart from the universal behavior for 
long orbits due to the violation of unitarity in a semiclassical 
approximation. The transition point from universal to non-universal 
statistics is distinctively different for the two parity subspaces in the 
Baker map. It is linked to the magnitude of the semiclassical error made when 
approximating the quantum density of states by semiclassical periodic orbit
formulae and is controlled by the largest deviation of a semiclassical
eigenvalue from the unit circle. 
This behaviour is probably generic for quantum maps. Statistical properties 
of semiclassical expressions are universal in the non-asymptotic regime, the 
transition point at which universality breaks down is, however, system 
dependent and controlled by the semiclassical error.
Imposing unitarity on a 
semiclassical approximation makes it possible to discard 
exponentially growing terms in the periodic orbit correlation 
function which in turn uncovers universal correlation even above the
transition point, as demonstrated here for the Baker map. 

The study presented here establishes for the first time unambiguously the 
existence of universal periodic orbit correlations in a classical chaotic
system whose quantum counterpart shows RMT-eigenvalue statistics; 
the limitations due to the semiclassical approximations are discussed in 
detail. This sets a proper framework for studying the {\em origin} of 
these classical correlations. A better understanding
of the interplay between classical periodic orbit action correlations and
unitarity might shed light on the existence of universality in quantum spectra 
in general.\\[1cm]

\noindent
{\large \bf Acknowledgments}\\

\noindent
I would like to thank Stephen Creagh, John Hannay and Jon Keating for 
stimulating 
discussions and the Isaac Newton Institute, Cambridge, for the hospitality 
during the workshop 'Disordered Systems and Quantum Chaos' where parts of 
this work have been carried out.

%        Referencen Baker Map      16.03.99
%

\end{document}